\documentclass[10pt]{iopart}

\usepackage[utf8]{inputenc}
\usepackage[T1]{fontenc}
\usepackage{amssymb,amsfonts,textcomp}
\usepackage{graphicx}
\usepackage[dvipsnames]{xcolor}
\usepackage{bm}
\usepackage{hyperref}

\newcommand{\Eqref}[1]{(\ref{#1})}
\newcommand{\Figref}[1]{figure~\ref{#1}}
\newcommand{\Secref}[1]{section~\ref{#1}}

\newcommand{\revision}[1]{\textcolor{black}{#1}}

\newcommand{\dipc}{Donostia International Physics Center (DIPC), E-20018, Donostia-San Sebasti\'an, Spain}
\newcommand{\iker}{IKERBASQUE, Basque Foundation for Science, E-48013, Bilbao, Spain}
\newcommand{\cfm}{Centro de F\'{\i}sica de Materiales (CFM) CSIC-UPV/EHU, E-20018, Donostia-San~Sebasti\'an, Spain}
\newcommand{\dtu}{Department of Physics, Technical University of Denmark, DK-2800 Kgs.~Lyngby, Denmark}
\newcommand{\dtucomp}{DTU Computing Center, Department of Applied Mathematics and Computer Science, Technical University of Denmark, DK-2800 Kgs.~Lyngby, Denmark}

\newcommand{\ie}{\emph{i.e.}}
\newcommand{\eg}{\emph{e.g.}}

\date{\today}

\begin{document}

\title{Mach--Zehnder-like interferometry with graphene nanoribbon networks}

\author{Sofia Sanz,$^1$ Nick Papior,$^2$ G\'eza Giedke,$^{1,3}$ Daniel S\'anchez-Portal,$^{4}$ Mads Brandbyge,$^{5}$ Thomas Frederiksen$^{1,3}$}

\address{$^1$ \dipc}
\address{$^2$ \dtucomp}
\address{$^3$ \iker}
\address{$^4$ \cfm}
\address{$^5$ \dtu}

\ead{sofia.sanz@dipc.org, thomas\_frederiksen@ehu.eus}

\begin{abstract}
We study theoretically electron interference in a Mach--Zehnder-like geometry formed by four zigzag graphene nanoribbons (ZGNRs) arranged in parallel pairs, one on top of the other, such that they form intersection angles of 60$^\circ$.
Depending on the interribbon separation, each intersection can be tuned to act either as an electron beam splitter or as a mirror, enabling tuneable circuitry with interfering pathways.
Based on the mean-field Hubbard model and Green's function techniques, we evaluate the electron transport properties of such 8-terminal devices and identify pairs of terminals that are subject to self-interference.
We further show that the scattering matrix formalism in the approximation of independent scattering at the four individual junctions provides accurate results as compared with the Green's function description, allowing for a simple interpretation of the interference process between two dominant pathways.
This enables us to characterize the device sensitivity to phase shifts from an external magnetic flux according to the Aharonov--Bohm effect as well as from small geometric variations in the two path lengths.
The proposed devices could find applications as magnetic field sensors and
as detectors of phase shifts induced by local scatterers on the different segments, such as adsorbates, impurities or defects.
The setup could also be used to create and study quantum entanglement. 
\end{abstract}

\maketitle
\ioptwocol

\section{Introduction}
Over the past decade the field of electron quantum optics, where electrons play the role of photons in quantum-optics like experiments, has witnessed strong theoretical and experimental advances.
For instance, several electronic analogues of optical setups have been implemented, 
such as the Mach--Zehnder \cite{Ji2003, Roulleau2007} and Fabry--P\'erot \cite{Zhang2009, McClure2009, Carbonell-Sanroma2017} interferometers, as well as the Hanbury Brown--Twiss \cite{Henny1999, Oliver1999, Samuelsson2004, Neder2007} geometry, enabling studies of fermion antibunching and the two-particle Aharonov--Bohm \cite{Splettstoesser2010} effect.

When it comes to electronic devices, graphene is an advantageous material
showing a high degree of quantum coherence even at moderately high temperatures \cite{CastroNeto2009}.
The similarities between electrons travelling ballistically in graphene constrictions and photons propagating in waveguides have placed the focus on this material for electron quantum optics applications.
For instance, the electron wave nature has manifested in refraction effects in $p$-$n$ junctions, \eg, when transmitted across a boundary separating regions of different doping levels \cite{RiMaLi.15.Snaketrajectoriesin,Chen2016}.

In particular, within the group of graphene derivatives and nanostructures, graphene nanoribbons (GNRs) offer attractive characteristics for electron quantum optics.
First, the confinement of electrons to one dimension (1D) provides a versatile, width-dependent electronic structure which can include the appearance of a band gap and spin-polarized edge-states \revision{as, \eg, in the case of GNRs of zigzag edge topology (ZGNRs)} \cite{Son2006a, Son2006b}.
Secondly, it has been experimentally demonstrated that GNRs possess long coherence lengths, that can reach values of the order of $\sim 100$ nm \cite{Minke2012, Baringhaus2014, Aprojanz2018}.
Furthermore, ballistic transport in \revision{Z}GNRs can be rather insensitive to edge defects because the current flows maximally through the center of the ribbon as a consequence of the dominating Dirac-like physics \cite{Zarbo2007}.

With respect to their experimental realization and feasibility, the emergence of bottom-up fabrication techniques has resulted in the fabrication of long, defect-free samples of GNRs via on-surface synthesis \cite{CaRuJa.10.Atomicallyprecisebottom, Ruffieux2016, ClOt.19.ControllingChemicalCoupling}.
This approach has also opened new possibilities to design $\pi$-magnetism in carbon nanostructures and to address localized, unpaired electron spins \cite{OtDiFr.22.Carbonbasednanostructures}.
Additionally, GNRs can also be picked up and manipulated with scanning tunneling probes \cite{Koch2012, Kawai2016, Wang2023}, suggesting the possibility of building two-dimensional multi-terminal GNR-based electronic circuits \cite{Areshkin2007, Jayasekera2007, Jiao2010, Cook2011, Botello-Mendez2011}.

One of the most elementary building blocks necessary to perform electron quantum optics experiments is the electron beam splitter, the electronic analog of a beam splitter for light, which coherently splits an incoming particle into a superposition of two states propagating in different output arms of the device.
Remarkably, it has been theoretically discovered that one GNR placed on top of another with an intersection angle close to $60^\circ$ enhances the electron transfer process between the ribbons, 
an effect related to the fact that the orientation of the honeycomb lattices of the bottom and top ribbons are aligned \cite{Brandimarte2017, Lima2016, Sanz2020}.
In fact, valence- or conduction-band electrons injected in such a four-terminal device are scattered into only two of the four possible outgoing directions without reflection.
Depending on the width of the GNR, interlayer separation, and energy of the traversing electrons, 
the branching ratio can be varied, resulting in different behaviours such as mirrors, beam splitters (half-transparent mirrors), and energy filters \cite{Sanz2020}.
Furthermore, \revision{the magnetic instabilities that appear due to the localization of the edge states in ZGNRs near the Fermi level} \cite{Fujita1996, Son2006a, Magda2014} make junctions of ZGNRs even more interesting since they can spin-polarize the transmitted electrons \cite{Sanz2022}.

With these fundamental components one can consider building a GNR-based Mach--Zehnder-like interferometer, which can be used for a variety of tasks from sensing of magnetic fluxes or local electric fields to measuring indistinguishability \cite{Neder2007}, statistics \cite{Law2006}, and coherence length \cite{Jo2022} of the charge carriers or generating entanglement between them \cite{Signal2005,Haack2011,Vyshnevyy2013}.
Other two-path setups have been demonstrated to act as a manipulable
flying qubit architecture \cite{Yamamoto2012} using the Aharonov--Bohm (AB) effect \cite{Aharonov1959}.
In graphene, the AB effect has been \revision{studied} in ring-shaped nanostructures both theoretically \cite{Recher2007, Schelter2012, Duca2015, Mrenca2016} and experimentally \cite{Russo2008}, \revision{and more recently also considered in bipolar hybrid monolayer-bilayer junctions \cite{Mirzakhani2023}}.
It has also been observed in a graphene quantum-Hall system for spin- and valley-polarized edge states \cite{Wei2017}.
However, these nanostructures are in general difficult to produce. Moreover, when it comes to electron interference, clean systems and long phase-coherence length are required.
Therefore, ZGNRs should provide an outstanding platform for electron quantum optics.

\begin{figure}
	\includegraphics[width=\columnwidth]{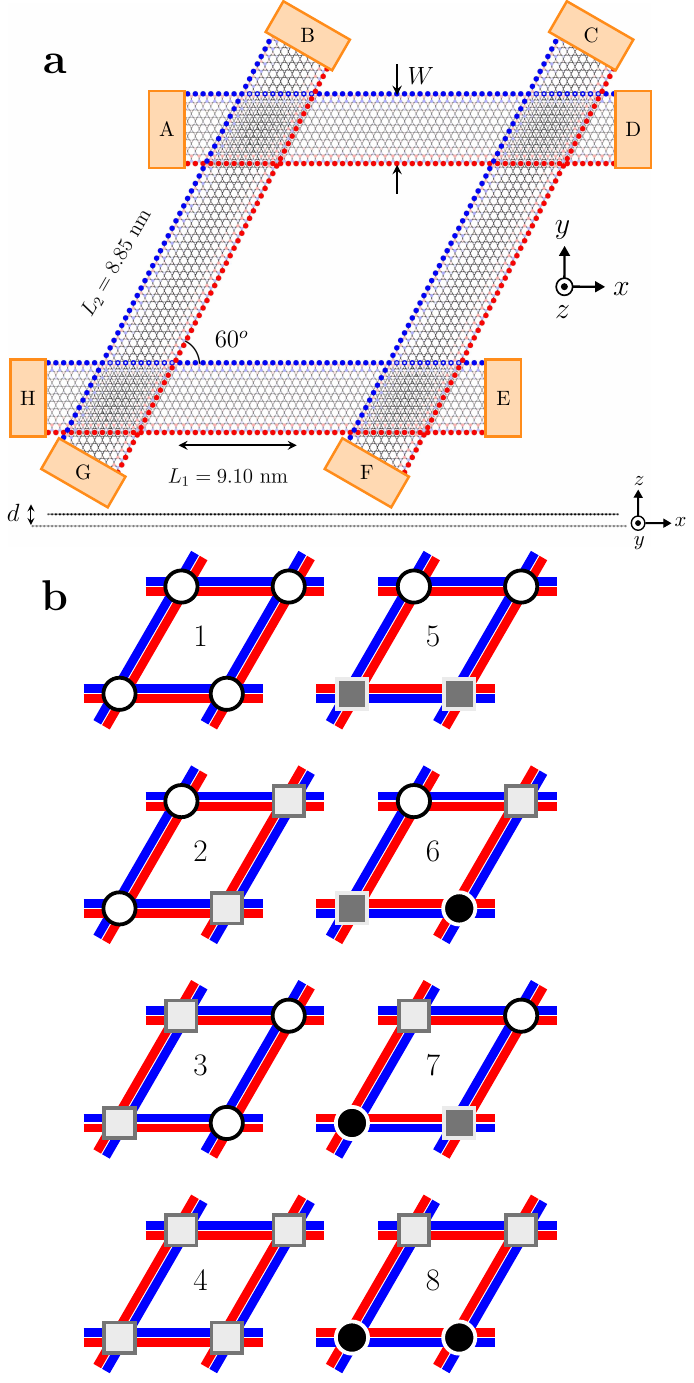}
	\caption{Illustration of the general setup. (a) Top view of the ZGNR interferometer. The eight terminals, labelled A-H, are represented as orange rectangles. The colored edges show the local spin polarization obtained as a solution to the MFH model with open boundary conditions. The width of the ZGNRs ($W=10$) is defined as the number of carbon atoms across the ribbon. $d$ stands for the vertical separation between the top and bottom ribbons while $L_{1,2}$ is the distance between the center of two junctions. (b) Representations of different low-energy spin configurations (labelled 1 to 8) differentiated by the edge polarization (red/blue for up/down spins) in each of the four ZGNRs. The symbols (white and black circles, and light and dark gray squares) represent the 4 different spin configurations of the individual junctions (each symbol represents a pair of different shadings that are related by spin inversion).}
	\label{fig:def-terminals}
\end{figure}

Here we propose a setup to study these phenomena which seem not too far away with the current experimental techniques.
Our interferometer, shown in \Figref{fig:def-terminals}(a), is formed by four crossed ZGNRs in a pairwise arrangement forming a parallelogram where the intersecting angle between the ribbons is $60^\circ$.
The ZGNR width is here selected to be of 10 carbon atoms ($W=10$); this choice is not particularly critical as qualitatively similar transport behavior is expected for other ribbon widths \cite{Sanz2020}.
We show that in the single-channel energy window near the Fermi level, electrons are transmitted essentially without reflection at each intersection.
This allows to describe the self-interference process by considering each junction as if they were independent scatterers for the incoming electrons.
Given the exclusive transmission into only two out of the four terminals in each beam splitter, the AB effect in the multiterminal setup does not suffer from complicated multi-path interferences that would lead to the loss of quantum information carried by coherent electrons reaching the other reservoirs.
This enables us to characterize the interferometer as a detector of phase shifts, \eg, induced by a transverse magnetic field, electric fields or any geometrical distortion or defect that changes the relative phase between the two available paths that will interfere.

\section{Methods}\label{sec:methods}

\subsection{Model Hamiltonian} 
The system, shown in \Figref{fig:def-terminals}(a), is divided into the device (scattering) region that contains the enclosed area by the crossing ribbons, and the eight semi-infinite ZGNRs (periodic electrodes), represented as orange rectangles.
The total Hamiltonian is correspondingly split into the different parts
$H_{T} =  H_{D} + \sum_{\alpha} (H_{\alpha} + H_{\alpha D})$, where $H_{D}$ is the device Hamiltonian, $H_{\alpha}$ the $\alpha$-electrode Hamiltonian, and $H_{\alpha D}$ the coupling between these two subsystems.

The effective Hamiltonian for the $\pi$-electrons, responsible for spin-polarization and transport phenomena, can be described in terms of the mean-field Hubbard (MFH) model with a single $p_z$ orbital per site \cite{OtDiFr.22.Carbonbasednanostructures, dipc_hubbard}, \ie,
\begin{eqnarray}\label{eq:TB-Hamiltonian}
	H_T &=& \sum_{i\sigma}\epsilon_{i}c^{\dagger}_{i\sigma}c_{i\sigma}^{\phantom{\dagger}} + \sum_{ij,\sigma} t_{ij}c^{\dagger}_{i\sigma}c_{j\sigma}^{\phantom{\dagger}}
	+ U\sum_{i,\sigma} n_{i\sigma}\left\langle n_{i\overline{\sigma}}\right\rangle,
\end{eqnarray}
where $c^{\dagger}_{i}$ ($c_{i}^{\phantom{\dagger}}$) creates (annihilates) an electron on site $i$ with spin $\sigma=\lbrace\uparrow, \downarrow \rbrace$, and $n_{i\sigma}=c^{\dagger}_{i,\sigma}c_{i\sigma}^{\phantom{\dagger}}$ is the number operator.
The Coulomb interaction is parametrized via the onsite repulsion $U$, which in the following is fixed to $U=3$ eV. The qualitative picture is not altered by its precise magnitude, only the quantitative results (like the size of the induced band gap).
Following Ref.~\cite{Sanz2020}, the matrix element $t_{ij}$ between orbitals $i$ and $j$ is described by Slater--Koster two-centre $\sigma$- and $\pi$-type integrals between two $p_{z}$ atomic orbitals \cite{Slater1954} as used previously for twisted bilayer graphene \cite{ TramblydeLaissardiere2010} and crossed GNRs \cite{Sanz2020, Sanz2022}.
\revision{We further fix the on-site energies $\epsilon_{i}=E_F$ equal to the Fermi energy $E_F$.
Given that ZGNRs develop a band gap for $U>0$ we define $E_F$ as the midgap value of the electrodes.}
As the junctions between the ribbons break translational invariance of the perfect ZGNRs, we use the nonequilibrium Green's functions (NEGF) \cite{Keldysh1965, Kadanoff1962} formalism to solve the Schr\"odinger equation for the open quantum system.
Details of the implemented MFH model with open boundary conditions \cite{dipc_hubbard} can be found in the supplemental material of Ref.~\cite{Sanz2022}.

Within the MFH approach, the self-consistent solution of a periodic ZGNR can be obtained by imposing one of the two possible symmetry-broken spin configurations at the edges.
This is, by fixing the $\uparrow$-spin majority at the lower edge of the ribbon and the $\downarrow$-spin majority at the upper edge, or vice versa.
While the ground state has zero net magnetic moment $m_z=0$ it displays antiferromagnetic order between unpaired spins at the edges.
For the device structure shown in \Figref{fig:def-terminals}(a) this implies in principle $2^8/2$
possible boundary conditions for the polarization of the electrode regions.
However, as shown previously, solutions with magnetic domain walls within the individual GNRs are energetically unfavorable compared to solutions with unaltered edge polarizations along the GNRs \cite{Sanz2022}.
This reduces the number of low-energy solutions to the 8 possibilities schematically shown in \Figref{fig:def-terminals}(b), with circles and squares representing two different magnetic orderings at a junction.
As an example, the calculated spin polarization for configuration 1 is superimposed on the structure in \Figref{fig:def-terminals}(a).

Each spin configuration for the total device also defines the spin configuration of the individual intersections between the ribbons.
For this reason, to show the different spin configurations we used different symbols to represent them as a combination of four different junctions.
There are two types of junctions, one that polarizes the outgoing beam, represented by a circle, and one that gives a non-polarized outgoing beam represented by a square\revision{, as shown in \cite{Sanz2022} and here in \Secref{sec:results}}.
The filling of the symbols (white and black, and light and dark gray) represent that one is the spin-inverted version of the other. 

\subsection{Peierls substitution method}\label{sec:Peierls}
To describe the system in the presence of a transverse magnetic field (\ie, along the $z$ direction) we use the Peierls substitution method \cite{Peierls1933}, where the gauge-invariance of the Schr\"odinger equation requires to transform the wave-function amplitude, or equivalently the hopping matrix elements as 
$t_{ij}\rightarrow t_{ij}e^{i\varphi_{ij}}$, 
where the phase shift
\begin{equation}\label{eq:peierls-phase}
    \varphi_{ij}=\frac{e}{\hbar}\int^{\mathbf{R}_{i}}_{\mathbf{R}_{j}} \mathbf{A}\cdot d\mathbf{r}
\end{equation} is the integral of the vector potential $\mathbf{A}$ along the \emph{hopping path}
with $\mathbf{R}_{k}=(x_{k},y_{k}, z_k)$ the coordinates of the orbital located at site $k$.

Given the relation between the magnetic field and the vector potential $\mathbf{B} = \nabla \times \mathbf{A}$, we have within the Landau gauge $\mathbf{A}(\mathbf{r}) = B_{0}x \, \hat{y}$, which leads to 
\begin{equation}\label{eq:peierls-phase-solved}
    \varphi_{ij}=\frac{\pi B_{0}}{2\Phi_{0}}(x_{j}-x_{i})\cdot(y_{j}+y_{i}),
\end{equation} where $\Phi_0=h/(2e)$ is the flux quantum.
We ignore the effect of a magnetic field outside the device region, \ie, the Peierls phases are not included in the leads.

We note that, while the ground state of ZGNRs display zero total magnetic moment $m_z=0$ (as mentioned above), the presence of a magnetic field $B$ can stabilize a high-spin configuration $|m_z|>0$ due to the Zeeman energy $\Delta E = g_S\mu_{B}m_z B$, where $g_S\approx 2$ is the electron spin $g$-factor and $\mu_B$ is the Bohr magneton.
Within our model calculations for 10-ZGNRs (Slater--Koster parametrization and $U=3$ eV), such a high-spin (excited-state) solution with ferromagnetic order across the spins at the GNR edges is obtained for $m_z=0.27$ per primitive cell.
The corresponding electronic energy is 5.7 meV/cell above the ground state, implying that a critical magnetic field of the order $B_c=182$ T is needed to make the two spin states degenerate.
In other words, as long as the magnetic field is below this critical value we expect the magnetic order of our device to be among those of \Figref{fig:def-terminals}(a), all corresponding to $m_z=0$.

\subsection{Electron transmission from Green's functions}
\label{sec:GreenFunctions}
To perform transport calculations we use the Green's function approach.
In particular, to obtain the transmission probabilities for each spin component $\sigma=\lbrace\uparrow,\downarrow\rbrace$ between leads $\alpha$ and $\beta$ ($T^{\sigma}_{\alpha\beta}$) we use the \cite{Landauer1957, Buettiker1985} for ballistic conductors,
\begin{equation}\label{eq:LandauerBuettiker}
    T^{\sigma}_{\alpha\beta}(E) = \mathrm{Tr}\left[\mathbf{\Gamma}^{\sigma}_{\alpha}\mathbf{G}^{\sigma}\mathbf{\Gamma}^{\sigma}_{\beta}\mathbf{G}^{\sigma\dagger}\right]
\end{equation}
where the retarded device Green's function is calculated as
\begin{equation}\label{eq:G-function}
    \mathbf{G}^{\sigma}(E) 
    = \Big[(E+i0^+)\, \mathbb{I} - \mathbf{H}^{\sigma}_{D} - \sum_{\alpha}\mathbf{\Sigma}^{\sigma}_{\alpha}\,\Big]^{-1},
\end{equation}
with $\mathbf{\Sigma}^{\sigma}_{\alpha}(E)$ being the retarded self-energy from $\alpha$, and
\begin{equation}
\mathbf{\Gamma}^{\sigma}_{\alpha} (E) = i\left(\mathbf{\Sigma}^{\sigma}_{\alpha} - \mathbf{\Sigma}^{\sigma \dagger}_{\alpha}\right)
\end{equation}
is the broadening matrix due to the coupling of the device region to lead $\alpha$.

The reflection probability can be conveniently written as a difference between the total number of open channels/modes available at that precise energy $M_{\alpha}^\sigma$ and the scattered transmission into all the $\beta\neq\alpha$ electrodes, \ie,
\begin{equation}
R_{\alpha}^\sigma(E) = M_{\alpha}^\sigma - \sum_{\beta\neq\alpha}T_{\alpha\beta}^\sigma.
\end{equation}

Computationally, we calculate the transmission probabilities from the Green's function using the open-source software TBtrans v4.1.5 \cite{Papior2017}.

\subsection{Scattering matrix formalism}
In order to analyze how electron transport is affected by scattering at each junction region we make use of the scattering matrix (S-matrix) approach, which can be easily computed from the retarded Green's function of the device for a given energy $E$ by means of the generalized Fisher-Lee relations \cite{Fisher1981,Papior2017}:
\begin{equation}
\label{eq:S-matrix}
	\mathbf{S}^{\sigma}_{\alpha\beta}(E) = - \delta_{\alpha\beta}\, \mathbb{I} 
	+ i \widetilde{\mathbf{\Gamma}}^{\sigma \dagger}_{\alpha}
	\mathbf{G}^{\sigma}
	\widetilde{\mathbf{\Gamma}}^{\sigma\phantom{\dagger}}_{\beta} \,,
\end{equation}
where 
\begin{equation}
\widetilde{\mathbf{\Gamma}}^{\sigma}_{\alpha}(E)
= \mathrm{diag}\lbrace\sqrt{\gamma^{\sigma}_{\alpha}}\rbrace \mathbf{U}^{\sigma}_{\alpha}
\end{equation}
is related to the level broadening matrix $\mathbf{\Gamma}^{\sigma}_{\alpha}$ by
\begin{equation}
 \mathbf{\Gamma}^{\sigma}(E) = \mathbf{U}^{\sigma\dagger}_{\alpha}\, \mathrm{diag}\lbrace \gamma^{\sigma}_{\alpha} \rbrace \mathbf{U}^{\sigma}_{\alpha}  \equiv  \widetilde{\mathbf{\Gamma}}^{\sigma \dagger}_{\alpha}\widetilde{\mathbf{\Gamma}}^{\sigma\phantom{\dagger}}_{\alpha}.
\end{equation}
Here $\mathbf{U}^{\sigma}_{\alpha}$ is the unitary matrix whose rows are the normalized eigenvectors of $\mathbf{\Gamma}^{\sigma}_{\alpha}$, which physically map into the transverse modes of the electrode that are coupled to the device by a strength given by the eigenvalues $\lbrace\gamma^{\sigma}_{\alpha}\rbrace$.
This enables a numerical simplification since one can discard the modes that are actually decoupled from the device, \ie, neglecting those vectors of $\mathbf{U}^{\sigma}_{\alpha}$ associated to $\gamma^{\sigma}_{\alpha}\simeq 0$.
We note here that the S-matrix in \Eqref{eq:S-matrix} for $N$-terminal devices is unitary as it can readily be verified that $\sum_\nu\mathbf{S}^{\sigma\dagger}_{\alpha\nu}\mathbf{S}^\sigma_{\beta\nu} = \delta_{\alpha\beta}\, \mathbb{I}$.

For $\alpha\neq \beta$ ($\alpha = \beta$) $\mathbf{S}^{\sigma}_{\alpha\beta}$ represents the transmission (reflection) amplitude matrix.
The corresponding transmission probability can be computed as
\begin{eqnarray}\label{eq:T-probability}
	T^{\sigma}_{\alpha\beta}(E) = \mathrm{Tr}\left[ \mathbf{S}^{\sigma\dagger}_{\alpha\beta}\mathbf{S}^{\sigma}_{\alpha\beta}\right],
\end{eqnarray}
where the trace runs over the transverse modes, recovering the Landauer-B\"uttiker conductance formula \cite{Landauer1957, Buettiker1985} written in \Eqref{eq:LandauerBuettiker}.

In the following we focus on the equilibrium situation where all leads have the same chemical potential ($\mu_{\alpha} = E_F =0$) and the temperature of the system is always $T=0$.
We modelled the Hamiltonians and obtained the self-energies $\mathbf{\Sigma}^{\sigma}_{\alpha}$ using recursion \cite{Sancho1985} as implemented in the open source, python-based SISL package \cite{Papior2017, zerothi_sisl}, while the scattering matrices are obtained using our own custom scripts.
For benchmarking our transport calculations based on the scattering-matrix formalism we also used the Green's function method for the whole device as explained in \Secref{sec:GreenFunctions}.

\section{Results}\label{sec:results}
\subsection{Independent-scatterers approximation}
The overall scattering matrix of the full device can be obtained by combining each junction's S-matrix coherently using the Feynman paths \cite{Cahay1988} to simulate the electronic path through the 8-terminal interferometer.
However, we know from previous results \cite{Sanz2020, Sanz2022} that electrons injected in ZGNR intersections are transmitted without reflection in the single-band energy region.
Under these circumstances the problem becomes significantly easier, and can be addressed by combining the S-matrices of each junction by using simple matrix multiplications, as the interference terms between the incoming and reflected beams are suppressed, leading to a computationally easier approach as compared to the full NEGF inversion.
\begin{figure}
\centering
	\includegraphics[width=\columnwidth]{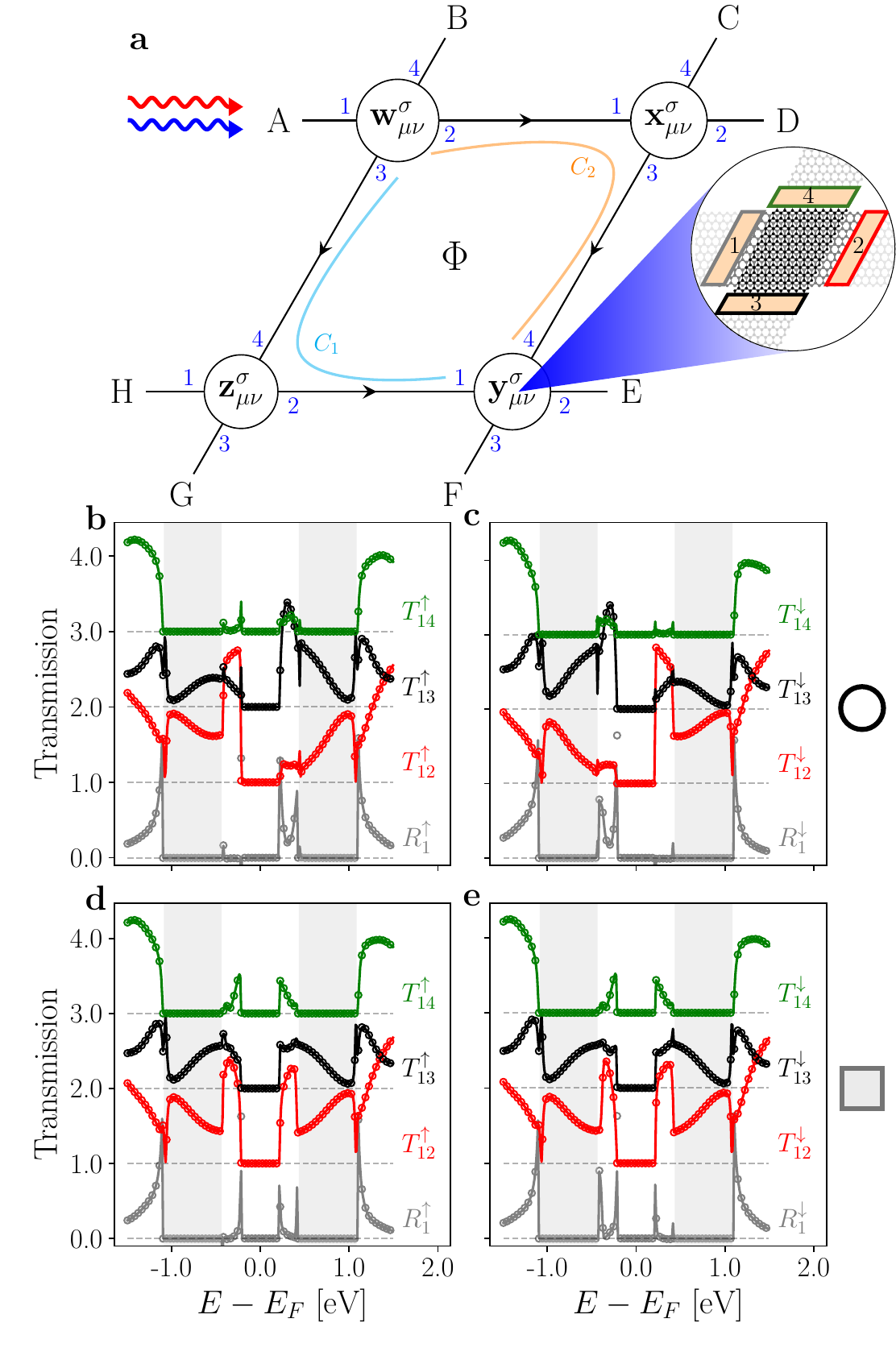}
	\caption{\revision{(a)} Sketch of the electronic circuit composed of four independent scatterers that conform the interferometer. Matrices $\mathbf{w, x, y, z}$ correspond to the S-matrix of the independent 4-terminal junction of two ZGNRs, as shown in the inset figure. Each junction can have in principle different scattering matrices, according to \Figref{fig:def-terminals}. The electrons are injected from terminal A (source). $\Phi$ denotes the magnetic flux that is created by the area enclosed within the interferometer. \revision{(b)-(c) Transmission ($T^{\sigma}_{\alpha\beta}$) (vertically offset) and reflection $R^{\sigma}_{\alpha}$ for $\sigma=\uparrow$ and $\sigma=\downarrow$, respectively, for terminals $\alpha=1$ and $\beta=2,3,4$ for a \emph{single} GNR crossing with the spin configuration represented by a circle in \Figref{fig:def-terminals}(b). (d)-(e) Same as panels b and c but for a single crossing with the spin configuration represented by a square in \Figref{fig:def-terminals}(b). The transmission functions were calculated with TBtrans \cite{Papior2017} (solid lines) and with the scattering matrix formalism (open circles). The single-channel energy region is shaded in light gray.}}
\label{fig:independent-scatt-sketch}
\end{figure}

In \Figref{fig:independent-scatt-sketch}\revision{(a)} we sketch the eight-terminal interferometer as a composition of four independent scatterers represented by $\mathbf{s}^\sigma_{\mu\nu}$ (lower-case), with $\mathbf{s}\in \lbrace \mathbf{w,x,y,z}\rbrace$, the S-matrix corresponding to each 4-terminal junction between two ZGNRs. Each junction $\mathbf{w,x,y,z}$  can, in principle, have different spin configurations, as shown in \Figref{fig:def-terminals}(b).
\revision{In \Figref{fig:independent-scatt-sketch}\revision{(b,c)} we show the transmission probabilities for a single crossing with spin configuration indicated with a circle in \Figref{fig:def-terminals}(b) for spin $\sigma=\uparrow,\downarrow$, respectively. Similarly, in \Figref{fig:independent-scatt-sketch}(d,e) we show these results for a crossing with spin configuration sketched with a square in \Figref{fig:def-terminals}(b). Here we can observe that, on the one hand, there is no transmission into port 4 in any of the crossings (green curves) and no reflection (gray curves) in the single channel energy region (shaded area). On the other hand, by looking at \Figref{fig:independent-scatt-sketch}(b-e) we can see that $T^{\sigma}_{12}$ and $T^{\sigma}_{13}$ are spin-dependent quantities (spin-polarizing beam splitter) for the crossing with spin configuration represented by a circle, while in the case of the crossing with spin configuration represented by a square, these transmission probabilities are spin independent (non-polarizing junction). For instance, by comparing \Figref{fig:independent-scatt-sketch}(b) and (c) it can be seen that $T^{\uparrow}_{\alpha\beta}(E)\approx T^{\downarrow}_{\alpha\beta}(-E)$ for $|E|\lesssim 1.1$ eV. For this system, we find that the polarizing crossing displays a maximum spin polarization in the transmission probability (within the single-channel energy region) of around $|T^{\uparrow}_{\alpha\beta}-T^{\downarrow}_{\alpha\beta}|\sim 0.45$ close to $E\sim 0.5$ eV, for $\alpha=1$ and $\beta=2,3$. Similar results were found for a crossing of slightly narrower GNRs (8-ZGNRs) in Rref.~\cite{Sanz2022}. Note that here we compare two methods to obtain the transmission probabilities, the scattering matrix formalism (open circles) and the NEGF method as implemented in TBtrans \cite{Papior2017} (solid lines), which give essentially the same result.}

The S-matrix of the complete interferometer, $\mathbf{S}^{\sigma}_{\alpha\beta}$ (uppercase), can then be written in terms of 
$\mathbf{s}^\sigma_{\mu\nu}$ with appropriate connection of in- and outgoing electrode indices ($\mu,\nu\in \lbrace 1,2,3,4 \rbrace$).
For electrons injected in the device from terminal A, one has:
\numparts
\begin{eqnarray}
	\mathbf{S}^{\sigma}_{\mathrm{AB}}
        &=& \mathbf{w}^{\sigma}_{14} \label{eq:s-matrices-approx-SAB}, \\
	\mathbf{S}^{\sigma}_{\mathrm{AC}}
        &=& \mathbf{w}^{\sigma}_{12} \mathbf{x}^{\sigma}_{14} \label{eq:s-matrices-approx-SAC}, \\
	\mathbf{S}^{\sigma}_{\mathrm{AD}}
        &=& \mathbf{w}^{\sigma}_{12} \mathbf{x}^{\sigma}_{12} \label{eq:s-matrices-approx-SAD}, \\ 
	\mathbf{S}^{\sigma}_{\mathrm{AE}}
        &=& \mathbf{w}^{\sigma}_{12} \mathbf{x}^{\sigma}_{13} \mathbf{y}^{\sigma}_{42} + \mathbf{w}^{\sigma}_{13} \mathbf{z}^{\sigma}_{42} \mathbf{y}^{\sigma}_{12} \label{eq:s-matrices-approx-SAE},\\
	\mathbf{S}^{\sigma}_{\mathrm{AF}}
        &=& \mathbf{w}^{\sigma}_{12} \mathbf{x}^{\sigma}_{13} \mathbf{y}^{\sigma}_{43} + \mathbf{w}^{\sigma}_{13} \mathbf{z}^{\sigma}_{42} \mathbf{y}^{\sigma}_{13}  \label{eq:s-matrices-approx-SAF}, \\
	\mathbf{S}^{\sigma}_{\mathrm{AG}}
        &=& \mathbf{w}^{\sigma}_{13} \mathbf{z}^{\sigma}_{43} \label{eq:s-matrices-approx-SAG}, \\
	\mathbf{S}^{\sigma}_{\mathrm{AH}}
        &=&  \mathbf{w}^{\sigma}_{13} \mathbf{z}^{\sigma}_{41} \label{eq:s-matrices-approx-SAH}.
\end{eqnarray}
\endnumparts

\begin{figure*}
	\centering
	\includegraphics[width=\textwidth]{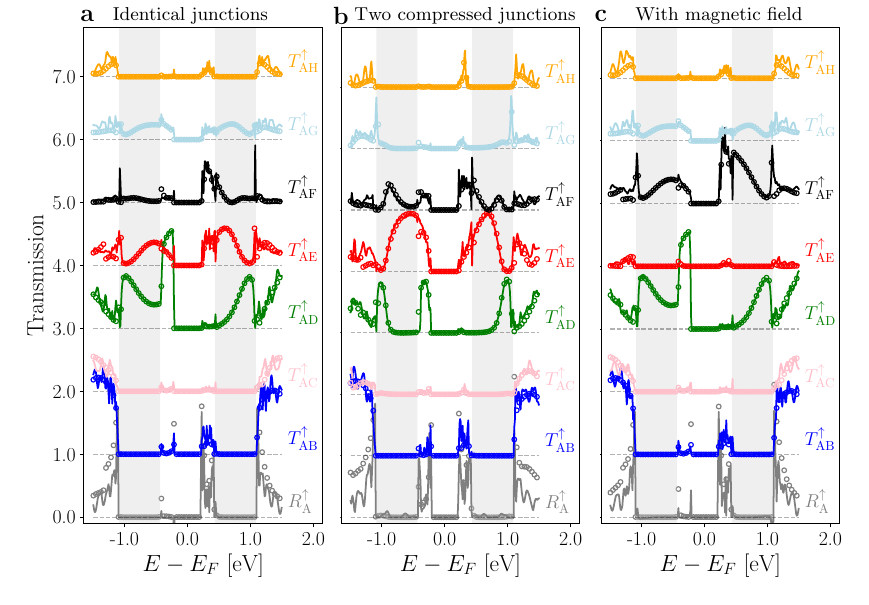}
 \vspace{-10mm}
	\caption{Transmission probabilities between the pairs of terminals of an 8-terminals interferometer built with 10-ZGNR and spin configuration 1 (see \Figref{fig:def-terminals}), with spin $\sigma=\uparrow$ obtained for the complete system with TBtrans \cite{Papior2017} (full lines) as well as with the independent-scatterers approximation (open circles).
    (a) Four identical junctions (interribbon distance $d = 3.34$ {\AA}).
    (b) Junctions \textbf{x} and \textbf{z} have a reduced interribbon distance $d = 3.19$ {\AA} relative to the junctions \textbf{w} and \textbf{y} ($d = 3.34$ {\AA}).
    (c) Four identical junctions ($d = 3.34$ {\AA}) in presence of a transversal magnetic flux $\Phi=\Phi_{0}$ within the interferometer.
    In all panels shaded areas indicate the (single subband) energy regions where there is only one available transmission channel.
	The curves are vertically offset by integer values. The horizontal dashed lines indicate $T^{\uparrow}_{\alpha\beta}=0$.
	}
	\label{fig:S-matrix-approx-transmission}
\end{figure*}

\begin{figure*}
	\centering
	\includegraphics[width=\textwidth]{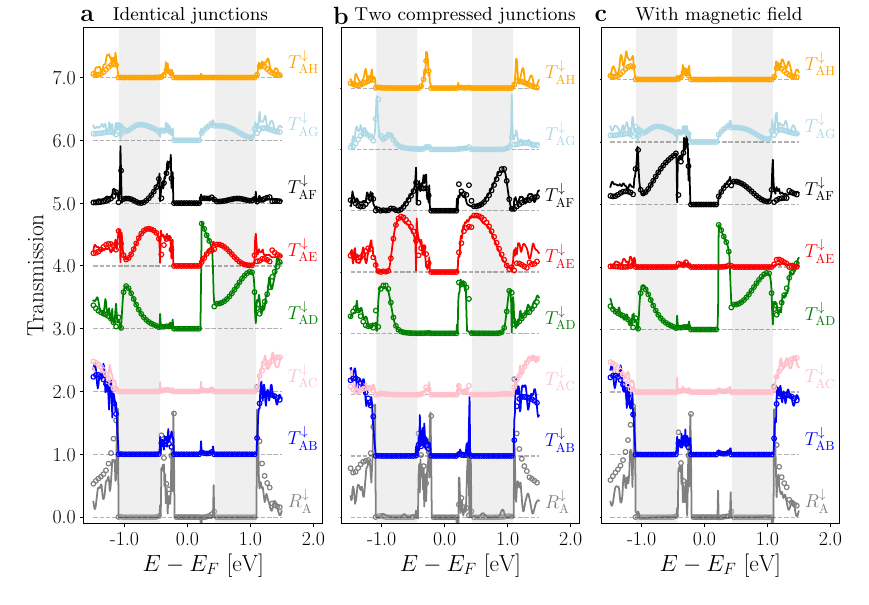}
 \vspace{-10mm}
	\caption{\revision{Same analysis as shown in \Figref{fig:S-matrix-approx-transmission} but for electrons with spin $\sigma=\downarrow$.}
	}
	\label{fig:S-matrix-approx-transmission-down}
\end{figure*}

To test our approximation, we compare in \Figref{fig:S-matrix-approx-transmission}(a) the transmission probabilities obtained with TBtrans \cite{Papior2017} (solid lines) for the whole interferometer, with those obtained with the independent-scatterers approximation (open circles) using \Eqref{eq:s-matrices-approx-SAB}-\Eqref{eq:s-matrices-approx-SAH}).
As shown in this panel, within the single-mode energy region (shaded areas), the approximation of independent scatterers yields a perfect overlap with the full solution.
This is due to the lack of reflection and/or interband scattering in the ZGNRs junctions for electrons in the single-mode energy region, \revision{as shown in \Figref{fig:independent-scatt-sketch}(b-e)}.
In fact, for larger energies, the approximate (open circles) and full solution (solid lines) start to deviate, since here one should also take into account the contribution of the reflected beams between independent scatterers to coherently combine the scattering matrices of the junctions.
\revision{For completeness, we performed the same analysis shown in \Figref{fig:S-matrix-approx-transmission} for electrons with the opposite spin $\sigma=\downarrow$, in \Figref{fig:S-matrix-approx-transmission-down}. It can be seen that the same observations listed above hold for the other spin component.}
%

\subsection{Deviations from standard Mach-Zehnder setup}
We note that in the ideal MZI interferometer two of the junctions should act as 50:50 electron beam splitters while the other two should act as perfect mirrors (where the outgoing electrons are transmitted with zero probability through those output ports).
In other words, the only non-zero transmission probabilities should be $T^{\sigma}_{AE}$ and $T^{\sigma}_{AF}$ with $T^{\sigma}_{AE}+T^{\sigma}_{AF}\approx 1$ (at the single-mode energy range).
While in \Figref{fig:S-matrix-approx-transmission}(a) we can observe that $T^{\sigma}_{AD}\neq 0$ and $T^{\sigma}_{AG}\neq 0$, one could achieve the ideal MZI interferometer by forcing junctions $\mathbf{x}$ and $\mathbf{z}$ (considering an electron incoming from terminal A) to work as mirrors instead of beam splitters.
One way to change the inter-/intra-ribbon transmission ratio (quantity that controls the performance of the junction) is by modifying the distance between the on-top and bottom ribbons (see \Figref{fig:def-terminals}) for such junctions.
For instance, if the vertical distance between the ribbons is reduced, the transmission between them can be enhanced up to values close to $100\%$ (condition for a perfect mirror where the electron beam is fully transferred between the ribbons) \cite{Sanz2020}.

In \Figref{fig:S-matrix-approx-transmission}(b) we calculated the transmission probabilities using our independent-scatterers approximation with \Eqref{eq:s-matrices-approx-SAB}-\Eqref{eq:s-matrices-approx-SAH}, where the scattering matrices $\mathbf{x}^{\sigma}_{\mu\nu}$ and $\mathbf{z}^{\sigma}_{\mu\nu}$ here correspond to a junction between two 10-ZGNRs with a relative distance that is $\sim 4.5\%$ reduced with respect to the junctions $\mathbf{w}$ and $\mathbf{y}$ ($d=3.19$ {\AA} versus $d=3.34$ {\AA}).
All junctions have the same spin configuration.
In this figure we can observe that there is an energy range (for $E \in (-0.5,-0.7)\cup (0.5,0.7)$, approximately) where all $T^{\uparrow}_{A\beta}\approx 0$ except for $T^{\uparrow}_{AE}$ and $T^{\uparrow}_{AF}$ (ideal MZI).

Note that even in this scenario where the geometry of the interferometer has been distorted, the reflection probability still remains close to zero in the single-mode energy region (gray curves in \Figref{fig:S-matrix-approx-transmission}(b)).

\subsection{Magnetic-field dependence of scattering matrices}
In presence of a magnetic field perpendicular to the interferometer plane (parallel to the $z$-axis in this case), as the crossed ZGNRs enclose an area, there will be a magnetic flux encompassed by the ribbons (represented by $\Phi$ in \Figref{fig:independent-scatt-sketch}).
Under these circumstances, the transmission probabilities between certain pairs of incoming/outgoing ports will be affected by the presence of the magnetic flux, as the electron injected can acquire a phase ($\Delta\varphi$) by following certain paths that surround a region with non-zero vector potential $\mathbf{A}$ enclosed by the interferometer,
\begin{eqnarray}\label{eq:phase-shift}
	\Delta\varphi 
	&=\frac{\pi}{\Phi_{0}}\oint_{C}\mathbf{A}\cdot d\mathbf{l} = \pi\frac{\Phi}{\Phi_{0}}.
\end{eqnarray}
Here $\Phi=B_{0}A$ is the flux of an external magnetic field $B_{0}$ through the area $A$ enclosed by the contour $C=C_1 + C_2$ (see \Figref{fig:independent-scatt-sketch}).
Because the global phase is arbitrary we can split the phase difference into contributions $\pm\Delta\varphi/2$ for the two pathways.

From \Eqref{eq:s-matrices-approx-SAB}-\Eqref{eq:s-matrices-approx-SAH} we observe that electrons incoming from terminal A 
only display interference for the pathways into terminals E and F, since here the S-matrices are built from a sum of two paths, as sketched in \Figref{fig:independent-scatt-sketch}.
To compute the transmission probabilities using the independent-scattering approximation, including the the additional phase contribution due to the presence of the magnetic field, we use the modified equations
\numparts
\begin{eqnarray}
	\mathbf{S}^{\sigma}_{\mathrm{AE}}
	&=& \mathbf{w}^{\sigma}_{12} \mathbf{x}^{\sigma}_{13} \mathbf{y}^{\sigma}_{42}e^{-i\Delta\varphi/2} + \mathbf{w}^{\sigma}_{13} \mathbf{z}^{\sigma}_{42} \mathbf{y}^{\sigma}_{12}e^{i\Delta\varphi/2}\label{eq:transmission-interferenceAE},\\
	\mathbf{S}^{\sigma}_{\mathrm{AF}}
	&=& \mathbf{w}^{\sigma}_{12} \mathbf{x}^{\sigma}_{13} \mathbf{y}^{\sigma}_{43}e^{-i\Delta\varphi/2} + \mathbf{w}^{\sigma}_{13} \mathbf{z}^{\sigma}_{42} \mathbf{y}^{\sigma}_{13}e^{i\Delta\varphi/2}\label{eq:transmission-interferenceAF}.
\end{eqnarray}
\endnumparts
The corresponding transmission probabilities between these terminals show a periodic dependency on the magnetic flux as a result of the interference term between the two possible paths.

We only show calculations for the spin configuration 1 because the self-interference patterns are very similar for all spin configurations shown in \Figref{fig:def-terminals}(b).
We note, however, that the spin polarization of the outgoing electron beam will depend on the chosen spin configuration.
For instance, the spin-polarizing effect is absent for configuration 4 in the case where the geometry is perfectly symmetrical.
Nevertheless, it is worth mentioning that away from the crossing angle of 60$^{\circ}$ (likely situation when building this geometry experimentally), each four-terminal junction generally polarizes the outgoing beam independently from the spin configuration \cite{Sanz2022}.

To test how our approximation works with the magnetic field, we compare in \Figref{fig:S-matrix-approx-transmission}(c) the transmission probabilities obtained with TBtrans \cite{Papior2017} (solid lines) for the whole interferometer, with those obtained with the independent-scatterers approximation (open circles) when $\Phi=\Phi_{0}$ for a device with equal junctions, where $d=3.4$~{\AA}.
We choose such magnetic flux since it leads to a phase difference of $\Delta\varphi=\pi$, and thus to a complete extinction in one arm.
The solid lines plotted in this panel were computed using the Peierls substitution (explained in  \Secref{sec:methods}) with a magnetic field of $B_0 = \Phi_0/A=29.6$ T for the device of  \Figref{fig:def-terminals}, of area $A=8.85 \times 9.10\times \sin(\pi/3)$ nm$^2$.

As shown in \Figref{fig:S-matrix-approx-transmission}, in the single-mode energy region (shaded areas), the approximation of independent scatterers (open circles) also yields a perfect overlap with the full solution (solid lines) in presence of a transverse magnetic field.

Another important result seen in \Figref{fig:S-matrix-approx-transmission} is that the reflection probability $R^{\uparrow}_{\mathrm{A}}$ is zero in the single-mode energy region both in presence or absence of a magnetic flux.
Moreover, we also observe, as predicted, that the transmission probabilities that do not involve two possible paths (\ie, $\textbf{S}^{\sigma}_{AB}, \ \textbf{S}^{\sigma}_{AC}, \ \textbf{S}^{\sigma}_{AD}, \ \textbf{S}^{\sigma}_{AG}, \ \textbf{S}^{\sigma}_{AH}$) are insensitive to the magnetic field since the only curves that are affected by $\Phi\neq 0$ are $T^{\sigma}_{AE}$ and $T^{\sigma}_{AF}$.

\begin{figure*}
    \centering
	\includegraphics[width=\textwidth]{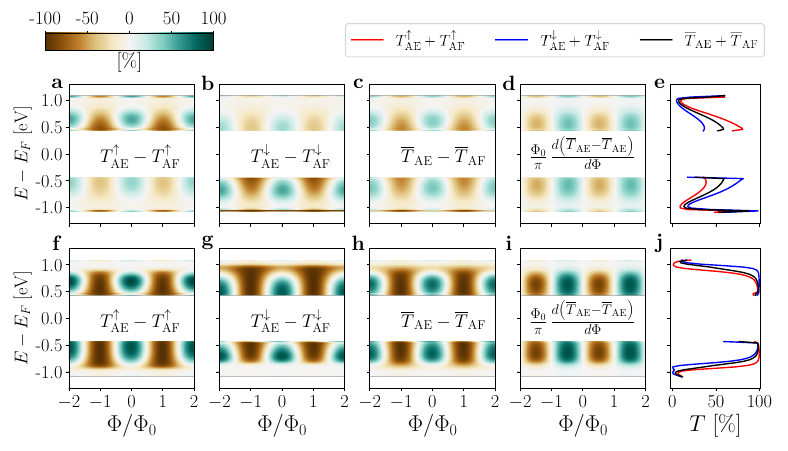}
	\caption{Interference pattern with the magnetic flux. Transmission probabilities as a function of the incoming electron energy $E-E_{F}$ and the magnetic flux $\Phi$ between terminals $\alpha=\mathrm{A}$ and $\beta=\mathrm{E,F}$.
    (a-c) Transmission probability difference between terminal pairs AE and AF for $\uparrow$- and $\downarrow$-electrons and the average between these two components, respectively, for the device with four equal junctions.
    (d) Difference between the derivative of the averaged transmission probabilities between terminals AE and AF with respect to the magnetic flux in units of $\Phi_0$ for the device of four equal junctions. 
    (e) Sum of the transmission probabilities between terminal pairs AE and AF for $\uparrow$- and $\downarrow$-electrons and the average between these two components, respectively, for the device with four equal junctions.
    (f-j) Similar to panels (a-e) but for the ideal MZI with the two junctions $\mathbf{x}$ and $\mathbf{z}$ compressed ($d = 3.19$ {\AA}) to provide effective mirrors, \ie, $\mathbf{x}^{\sigma}_{13}= 1$ and $\mathbf{z}^{\sigma}_{42}= 1$. 
    Only single-channel energy regions are shown.}
	\label{fig:interference-patterns2}
\end{figure*}
\begin{figure*}
	\centering
	\includegraphics[width=\textwidth]{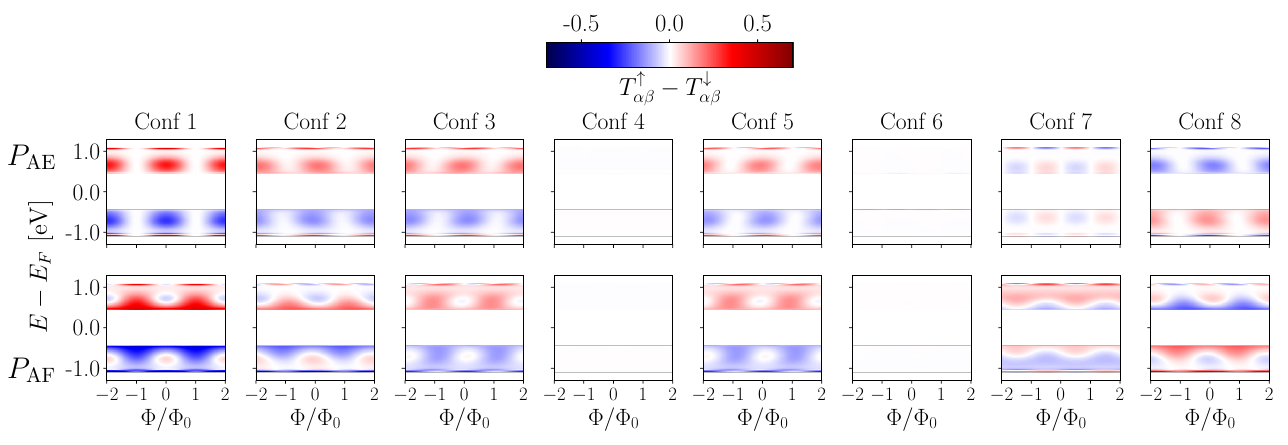}
	\caption{Spin polarization of the transmission probability as a function of the incoming electron energy $E-E_{F}$ and the magnetic flux $\Phi$ between terminals $\alpha=\mathrm{A}$ and $\beta=\mathrm{E,F}$ (top and bottom rows). Only single-channel energy regions are shown.}
	\label{fig:polarizations}
\end{figure*}
%

\subsection{Self-interference pattern with magnetic field}

In \Figref{fig:interference-patterns2} we show the interference patterns per spin channel of an interferometer by computing the transmission probability as a function of the incoming electron energy only in the single-mode energy window and for a magnetic flux $\Phi/\Phi_{0}\in [-2,2]$.
Only the single-band energy region is shown for which the independent-scatterers approximation provides accurate results.
We plot the difference $T^{\sigma}_{AE}-T^{\sigma}_{AF}$ for $\sigma={\uparrow,\downarrow}$ in \Figref{fig:interference-patterns2}(a,b) and its average in \Figref{fig:interference-patterns2}(c). 
In \Figref{fig:interference-patterns2}(d) we plot the difference between the derivatives of the transmission probabilities $\overline{T}_{AE}$ and $\overline{T}_{AF}$ with respect to the magnetic flux.
In this panel we can identify the regions of high sensitivity of the device.
In \Figref{fig:interference-patterns2}(e) we plot the sum of the transmission probabilities $T^{\sigma}_{AE}+T^{\sigma}_{AF}$ for $\sigma={\uparrow,\downarrow}$ and its average per spin channel.
While \Figref{fig:interference-patterns2}(a-e) correspond to a device formed of four equal junctions,
\Figref{fig:interference-patterns2}(f-j) show a similar analysis for the ideal MZI where junctions $\mathbf{x}$ and $\mathbf{z}$ are compressed ($d = 3.19$ {\AA}) to provide effective mirrors.
The figure clearly reveals the AB effect for electrons after self-interfering in the outgoing terminals.
We also see that the transmission probability is highly dependent on the electron energy and magnetic flux.
These transmission probabilities also display a slight dependency on the spin index of the electrons, as the junctions in configuration 1 polarize the outgoing electron beam \cite{Sanz2022}. 
By comparing \Figref{fig:interference-patterns2}(a-e) to \Figref{fig:interference-patterns2}(f-j) we observe that the device with compressed junctions acts as an ideal MZI where the electron beam is transmitted only into ports E and F without losses, as the corresponding transmission probabilities in this case reach values close to 100\%, while in the case of four equal junctions $T^{\sigma}_{AE}$ and $T^{\sigma}_{AF}$ reach maximum values of $\sim 50$\%.
\revision{Note that the sum of the transmission probabilities shown in \Figref{fig:interference-patterns2}(e,j) is constant with respect to the magnetic flux, since the magnetic field only modifies the relative phase acquired by an electron between the two paths that interfere, but this phase does not change the transmission probabilities between electrode A and electrodes B, C, D, G, and H.
Since the current must be conserved, while $T^{\sigma}_{AE}$ and $T^{\sigma}_{AF}$ are individually affected by the presence of the magnetic field, the sum of them must remain constant.}
As shown in \Figref{fig:S-matrix-approx-transmission}(c), the complete extinction of the transmission into one arm is independent of the transmission ratio between the ZGNRs of junctions \textbf{w} and \textbf{y} (as long as they are identical), while the contrast of the signal shown in \Figref{fig:interference-patterns2} is highly dependent on this ratio (and is optimal for 50:50 beam splitters).

For completeness, in \Figref{fig:polarizations} we show the degree of transport spin polarization for the 8 possible spin configurations of \Figref{fig:def-terminals}(b), defined as
\begin{eqnarray}\label{eq:polarization}
    P_{\alpha\beta} = T_{\alpha\beta}^{\uparrow}-T_{\alpha\beta}^{\downarrow}.
\end{eqnarray}
We observe that the device's spin configuration is not particularly relevant for the interference pattern, as the interference is determined by a periodic dependency on  the magnetic flux.
However, the spin configuration is important for controlling the spin polarization of the outgoing electron beam.
Interestingly, we observe that the spin polarization not only varies with the electron energy, but can actually be tuned with the magnetic field. But we also see that there are certain spin configurations that do not give a polarized outgoing beam (such as configurations 4 and 6).
In the case of configuration 4, it is easy to understand that the outgoing beam is not spin polarized as the four junctions in this case are non-polarizing \cite{Sanz2022}.
In the case of configuration 6, while there are two spin-polarizing junctions (\textbf{x} and \textbf{z}), the one in the outgoing junction \textbf{y} is the spin-inverted version of the one in the incoming junction \textbf{w}.
The multiplication of the corresponding scattering matrices results in a non-polarized outgoing beam.

\subsection{Other applications}
The MZI can be used to detect differential phase shifts between the two arms of the MZI, that could be caused, \eg, by defects, potentials in one of the four arms, path length, or by the charging state of nearby defects.

The functioning of the MZI depends crucially on the phase coherence of the electronic wave functions traveling along the two paths.
Thus, it is natural that the MZI can also be used to detect decoherence and measure, \eg, the electron's coherence length and how it is influenced by parameters such as temperature \cite{Seelig2001,Haack2011,Jo2022}.
Furthermore, the MZI can be used to learn about properties of the carriers producing the signal, \eg, to measure the degree of indistinguishability of electrons \cite{Samuelsson2004,Neder2007}, or the statistics of the charge carriers in the device \cite{Law2006}.

There are also several applications related to quantum information processing.
The MZI can be used to implement single-qubit rotations (on charge qubits, or with spin-dependent phases induced, \eg, by spin-orbit interaction also for spin qubits), or to perform entangling operations such as parity measurements \cite{Haack2010} or probabilistic entanglement generation \cite{Signal2005}. 

Since the consideration of electron-electron interaction, spin-orbit interaction and decoherence is beyond the scope of this article, we will illustrate the use of the proposed GNR-MZI for phase detection.

\begin{figure}
\centering
	\includegraphics[width=\columnwidth]{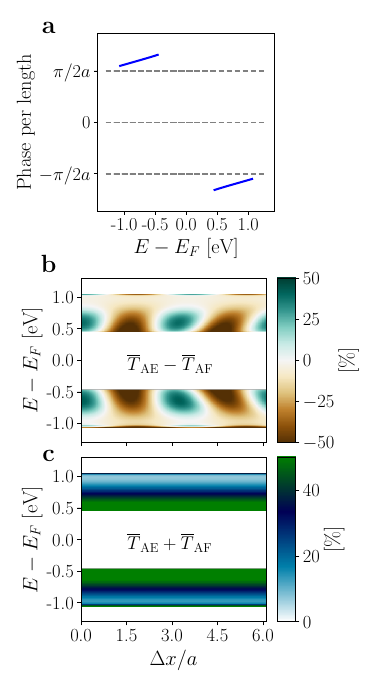}
	\caption{Sensor of geometrical distortions. (a) Phase shift per length as a function of electron energy calculated only for the single-mode energy range where $a=2.46$ {\AA} is the graphene lattice constant (size of the ZGNR unit cell along the periodic direction).
    (b) Difference and (c) sum of the average transmission probability per spin channel between ports A and E, F of a 10-ZGNR interferometer as a function of the relative difference in length between $C_1$ and $C_2$, calculated with the independent-scatterers approximation. We consider the interferometer in spin configuration 1.}
	\label{fig:phase-shift}
\end{figure}
As an example, here we consider the case of a geometrical distortion where the length of one of the paths \revision{($C_2$)} is longer than the other \revision{($C_1$)} that could be caused, \eg, by the presence of a fold in one of the ribbons composing it or a corrugation in the substrate underneath.
To simulate the presence of such geometrical distortion we include a section of a perfect ribbon between two ports to simulate the extra distance between the two paths.
We do not consider any modified hopping or on-site energy terms in the Hamiltonian, as a first approximation.

Electrons propagating through the perfect ribbon section are transmitted with 100\% probability.
However, they acquire a phase that depends on the size of such section, which can be determined by computing the complex part of the transmission amplitude (scattering matrix) of this system.
In the single-mode energy region this phase is easily determined since the scattering matrix of this section is of size $(1\times 1)$.
For this reason we can compute the overall scattering matrix of the interferometer by using the modified \Eqref{eq:s-matrices-approx-SAE} and \Eqref{eq:s-matrices-approx-SAF}:
\numparts
\begin{eqnarray}
	\tilde{\mathbf{S}}^{\sigma}_{\mathrm{AE}}
	&=& \mathbf{w}^{\sigma}_{12} \mathbf{x}^{\sigma}_{13} \mathbf{y}^{\sigma}_{42} e^{-i\chi \Delta x}
	+ \mathbf{w}^{\sigma}_{13} \mathbf{z}^{\sigma}_{42} \mathbf{y}^{\sigma}_{12} \label{eq:s-matrices-approx-SAE2},\\
	\tilde{\mathbf{S}}^{\sigma}_{\mathrm{AF}}
	&=& \mathbf{w}^{\sigma}_{12} \mathbf{x}^{\sigma}_{13} \mathbf{y}^{\sigma}_{43}e^{-i\chi \Delta x} + \mathbf{w}^{\sigma}_{13} \mathbf{z}^{\sigma}_{42} \mathbf{y}^{\sigma}_{13}  \label{eq:s-matrices-approx-SAF2},
\end{eqnarray}
\endnumparts
where, $\chi$ is the electronic phase shift per length unit of a perfect 10-ZGNR section, and $\Delta x$ represents the relative length difference between $C_1$ and $C_2$.
For this example we assume a longer distance between ports $\mathbf{w}$ and $\mathbf{x}$, making the $C_2$ path $\Delta x$ {\AA} longer than \revision{$C_1$}. Although other S-matrix elements of \Eqref{eq:s-matrices-approx-SAB}-\Eqref{eq:s-matrices-approx-SAH} are modified as well by the presence of such geometrical distortions (as, \eg, \Eqref{eq:s-matrices-approx-SAC} and \Eqref{eq:s-matrices-approx-SAD} considering that the section of a perfect 10-ZGNR is situated between junctions $\mathbf{w}$ and $\mathbf{x}$), the transmission probabilities associated to those matrices are not affected since a global phase does not change these values.
However, in the case of \Eqref{eq:s-matrices-approx-SAE}-\Eqref{eq:s-matrices-approx-SAF}, the presence of the 10-ZGNR section adds a relative phase between the two paths which affects the overall transmission probabilities associated to those S-matrices.

In \Figref{fig:phase-shift}(a) we plot the phase-shift per unit length acquired by an electron passing through a section of a perfect 10-ZGNR within the single-mode energy window.
In \Figref{fig:phase-shift}(b,c) we plot difference and sum, respectively, of the average transmission probabilities $\overline{T}_{AE}$ and $\overline{T}_{AF}$ as a function of the relative length difference between $C_1$ and $C_2$ ($\Delta x$), and the electron energy (only shown the single-mode energy range) in absence of an external magnetic field.
Here it is easy to see that the interference pattern will also depend on these kind of geometrical distortions that affect the relative phase acquired by an electron after travelling through paths $C_1$ and $C_2$.

\section{Conclusions}
In this work we have studied the performance of a device formed of four crossed ZGNRs as an  \emph{electron and spin} interferometer.
As ZGNRs host spin-polarized states due to the presence of electron interactions, we use the mean-field Hubbard Hamiltonian to describe the spin physics in this device by including the Coulomb repulsion term.
To solve the Schr\"odinger equation in each iteration step we use the NEGF formalism for this open quantum system.
Since the junctions create spin-polarizing scattering potentials \cite{Sanz2022}, the resulting transmitted electrons in the different outgoing directions are spin polarized as well.

Furthermore, since electrons are transmitted without reflection, we can consider the system as an array of independent scatterers by using the S-matrix of each 4-terminal junction and combining them correspondingly to compute the overall S-matrix for the full device.
The agreement between this approximation and the full solution is practically exact in the single-channel energy region, where the backscattering and transmission into the other output are zero.

Since some of the output ports can be reached following two possible paths, the transmission probability into these depends on the self-interference of the propagating electron.
Moreover, the self-interference pattern can be further tuned by applying an external uniform magnetic field as a consequence of the Aharonov--Bohm effect.
Interestingly, the self-interference pattern will depend not only on the electron energy and magnetic flux, but also on the spin index of the transmitted electrons.
To further analyze this effect, we also calculated the spin polarization in the transmission probability of the two outgoing directions of interest, as a function of the electron energy and magnetic flux.
While in the case of the interference pattern, the spin configuration is not particularly relevant, as the interference is dominantly determined by the cosine dependency of the magnetic flux, the spin configuration will determine the spin polarization of the outgoing electron beam in the possible outgoing ports.
For instance, depending on the combination of the spin configurations of the junctions, the resulting outgoing beam will be polarized or non-polarized.
Remarkably, the spin polarization not only varies with the electron energy, it can also be tuned by the existing magnetic flux.

Since the invention and further development of the single-electron source \cite{Feve2007}, performing electron-quantum-optics experiments at the single-particle level is now possible, where both emission and detection achieve efficiencies that reach values even larger than photon-based sources \cite{Edlbauer2022}.
While the major obstacle for quantum implementations with single
flying electrons is decoherence, here we propose a graphene-based interferometer in which spin-orbit and hyperfine interaction---the two major intrinsic sources of spin relaxation and decoherence in GaAs devices---are expected to be small due to carbon's low atomic mass and low abundance of spinful nuclei.
In fact, realizing a MZI in GNR-based nanostructures would set the stage for electron quantum optics experiments in this platform and provide evidence for its viability as a basis for quantum computing.

\section*{Acknowledgements}
We acknowledge fruitful discussions about the scattering matrix formalism with Pedro Brandimarte and Alexandre Reily Rocha.
This work was funded by the Spanish MCIN/AEI/ 10.13039/501100011033 (PID2020-115406GB-I00 and PID2019-107338RB-C66), the Basque Department of Education (PIBA-2020-1-0014), the University of the Basque Country (UPV/EHU) through Grant IT-1569-22, and the European Union’s Horizon 2020 (FET-Open project SPRING Grant No.~863098).

\section*{References}
\bibliographystyle{iopart-num-tf}
\providecommand{\newblock}{}

\end{document}